# Bayesian Optimised Collection Strategies for Fatigue Strength Testing

CM Magazzeni[1] *, R Rose[1] *, C Gearhart[2], J Gong[1], AJ Wilkinson[1]

[1] Department of Materials, University of Oxford, Parks Road, OX1 3PH, Oxford, United Kingdom

[2] National Renewables Energy Labs, Denver W Pkwy, Golden, CO 80401, United States

* : Corresponding author: christopher.magazzeni@queens.ox.ac.uk

## ABSTRACT

A statistical framework is presented enabling optimal sampling and analysis of constant life fatigue data. Protocols using Bayesian maximum entropy sampling are built based on conventional staircase and stress step methods, reducing the requirement of prior knowledge for data collection. The Bayesian Staircase method shows improved parameter estimation efficiency, and the Bayesian Stress Step method shows equal accuracy to the standard method at larger step size allowing experimentalists to lessen concerns of loading history. Statistical methods for determining model suitability are shown, highlighting the influence of protocol. Experimental validation is performed, showing the applicability of the methods in laboratory testing.

*Table 1: Glossary of symbols*

| Nomenclature | |
|---|---|
| $S$ | Stress amplitude |
| $\theta$ | Vector of model parameters |
| $\mu$ | Mean |
| $\sigma$ | Standard Deviation |
| $\Phi$ | Cumulative distribution function |
| $p$ | Probability |
| $\boldsymbol{x}$ | Vector of previous test stresses |
| $\boldsymbol{j}$ | Vector of previous test results |
| $x_{i,k}$ | Test stress of specimen $i$ at stress increment $k$ |
| $j_{i,k}$ | Test result of specimen $i$ at stress increment $k$ |
| $f_S$ | Probability density function of fatigue strength |
| $F_S$ | Cumulative distribution function of fatigue strength |
| $I$ | Shannon Information |
| $ent$ | Expected information (Entropy) |
| $\Upsilon$ | Gain in entropy |
| $\alpha, \beta, c$ | Basquin Constants |
| $U$ | Utility function |
| $M$ | Probability model |
| $L$ | Likelihood function |
| $\hat{L}$ | Maximum value of the likelihood function |
| $k$ | Number of parameters in the probability model |
| $n$ | Number of observations |
| **Abbreviations** | |
| HCF | High Cycle Fatigue |
| LCF | Low Cycle Fatigue |
| MAP | Maximum a Posteriori Probability |
| HPD | Highest Posterior Density |
| BIC | Bayes Information Criterion |
| CDF | Cumulative Distribution Function |
| PDF | Probability Density Function |

# 1 INTRODUCTION

The fatigue of materials affects the longevity of engineered components, causing failures in service with often-costly outcomes [1]. The life of materials is known to decrease with increasing cyclic plastic strain and/or stress. Populating the characteristic stress vs life—or SN curve—can be carried out by subjecting test specimens to cyclic loading. Fatigue data is often collected by considering the number of cycles for each specimen to fail "life testing", though sometimes – such as in the High Cycle Fatigue (HCF) region – the fatigue strength (distribution in strength at a given life) is needed. Obtaining better estimates of fatigue strength parameters could inform more efficient engineering designs, and faster data collection enables shorter, lower cost, development cycles. The staircase and step methods are the most common protocols for obtaining these results with single tests, or with multiple tests on an individual specimen, respectively. These protocols are described in brief in Table 2, with more extensive descriptions in papers and standards [2–8].

*Table 2: Description of conventional fatigue strength protocols.*

| Protocol | Description | Class |
|---|---|---|
| Staircase / Up-down | The first specimen is tested at the starting stress. For the next specimen, the stress is decreased or increased by a step size if the preceding specimen failed or ran out, respectively. The final state (failed or runout) of each specimen is recorded. | Single Test |
| Step | Specimens are initially tested at the starting stress. If the specimen runs out, it is tested again increasing the stress by the step size until failure. The runout and failure stress is recorded for each specimen. | Multiple Test |

In single-test strategies such as the staircase method, protocol parameters and data analysis are ambiguous, conflicting, or require prior knowledge of the system. As reported by Lin, Lu and Lee [9], there is disagreement within the standards on key protocol parameters for step testing such as sample size, starting stress, and step size as well as the method for analysing test data. The Dixon-Mood [3] method, a common analysis technique used in industry for staircase data, discards half of the test data (failures or runouts), assumes a normal distribution in failure life, and recommends stress increments approximately equal to the sample set standard deviation which is not initially known. Pollak's investigation [10] on simulated test data showed the effect of sample size and starting stress on estimates of the standard deviation and developed correction functions to increase the speed of convergence of the mean estimate. Engler-Pinto *et al*. developed a maximum likelihood-based framework for analysing staircase data which includes the information from all the specimens, further incorporating life testing to generate parametric probabilistic SN curves [11,12]. These studies all highlight the sensitivity of the protocol parameters in the analysis of single-test results.

In comparison to staircase protocols, stress step methods have the advantage that all specimens are tested to failure. However, there is significant debate over the reliability of step test data due to concerns over the effect of previous tests, through damage or coaxing [2,13–17]. Furthermore, as each specimen is tested multiple times, the experimental time taken per specimen can be higher than staircase testing. Nevertheless, these effects have not been fully characterised and the testing strategy remains in use across the academic literature [18,19,28–37,20–27]. For these reasons, step sizes are generally advised to be above 5% of the fatigue limit [4]. However, this requires prior knowledge of material properties and is dependent on the life being tested, as well as the continued debate over whether this is sufficient to avoid damage/coaxing. While larger step sizes reduce the effect of loading history, there is a trade-off between accuracy and precision: the details of sample distribution are lost when carrying out tests at larger step size.

There is also disagreement in the analysis of these results: the ASTM guide suggests using the mean of the runout and failure stress as the fatigue strength of that specimen [4], whereas Lerch *et al.* use the runout stresses only [17]. While some analysis methods aiming to account for coaxing or damage effects are available [30,33], extensive data are required to conclude their applicability and all agree that a larger step size reduces the effect.

We present two improved protocols for the collection of characteristic fatigue strength distribution parameters using a Bayesian Optimised Maximum Entropy Sampling method. Based on Lindley's approach [38–40], this broadly aims to design an experiment that maximises some expected experimental aim, such as the information gained from each subsequent specimen. In this field, a function is created which expresses the utility of testing at any location given the information from a prior. Bayesian optimal experimental design is discussed in reviews from DasGupta [41], Chaloner and Verdinelli [42], and more recently by Ryan, Drovandi *et al* [43,44]. We present this method applied to simulated results followed by its application in a laboratory experiment. The toolbox is available open source on the GitHub repository "BayesOptFatigue" [45].

## 2 APPLYING BAYES THEORY TO FATIGUE TESTING

The experimental design is optimised such that, based on information from all previous specimens, the subsequent specimen provides a maximised increase in the predictive capability of the model. To describe this, the $i^{th}$ specimen is tested at a stress level $x_i$ with result $j_i \in \{1,0\}$, indicating that the specimen has runout ($j_i = 0$) or failed ($j_i = 1$) by the target life. In tests with multiple stresses per specimen we denote the stress level by a second index $k$: $x_{i,k}$. As we consider only step testing protocols with constant step size, the test levels denoted by $k$ start from $x_{i,1}$ and incrementing by the step size $\xi$ such that for some specimen $x_i$:

$$x_{i,k} = x_{i,k-1} + \xi \qquad (1)$$

The data from all tested specimens is collected to form the matrices $\boldsymbol{x}$ and $\boldsymbol{j}$, which contain the stresses and results from every specimen and step increment such that every row corresponds to a specimen. For single test protocols, this matrix becomes a vector where each entry is a new specimen. The distribution in failure stress is modelled with parameters $\boldsymbol{\theta}$. For demonstration purposes, we start with the example of a normal distribution which is represented by the parameters mean ($\mu$) and standard deviation ($\sigma$), i.e. $\boldsymbol{\theta} = [\mu, \sigma]$.

A Bayesian approach evaluates the probabilities of model parameters given observations, built on Bayes rule Eq. (2):

$$p(\boldsymbol{\theta}|[\boldsymbol{j},\boldsymbol{x}]) = \frac{p(\boldsymbol{j}|[\boldsymbol{x},\boldsymbol{\theta}])p(\boldsymbol{\theta})}{p(\boldsymbol{j},\boldsymbol{x})} \qquad (2)$$

The posterior $p(\boldsymbol{\theta}|[\boldsymbol{j},\boldsymbol{x}])$ describes the probability supporting the model parameters. This posterior is inferred from the prior, $p(\boldsymbol{\theta})$, which describes the probability supporting the model without any data, the likelihood function $p(j|[x,\theta])$ which gives the likelihood of collecting the data given the model parameters, and finally the marginal likelihood $p(j,x)$ of collecting the data Eq. (3):

$$p(j,x) = \int d\theta p(j|[x,\theta])p(\theta) \qquad (3)$$

The prior used is constant with a range in mean and standard deviation that should cover a parameter space to confidently include the model parameters. A reasonable assumption in a metallic case could, for example, cover a mean from 0 MPa to the Ultimate Tensile Stress (UTS) and standard deviation between 1 and half the UTS. A discussion of initial prior choice is given in Section 8.1

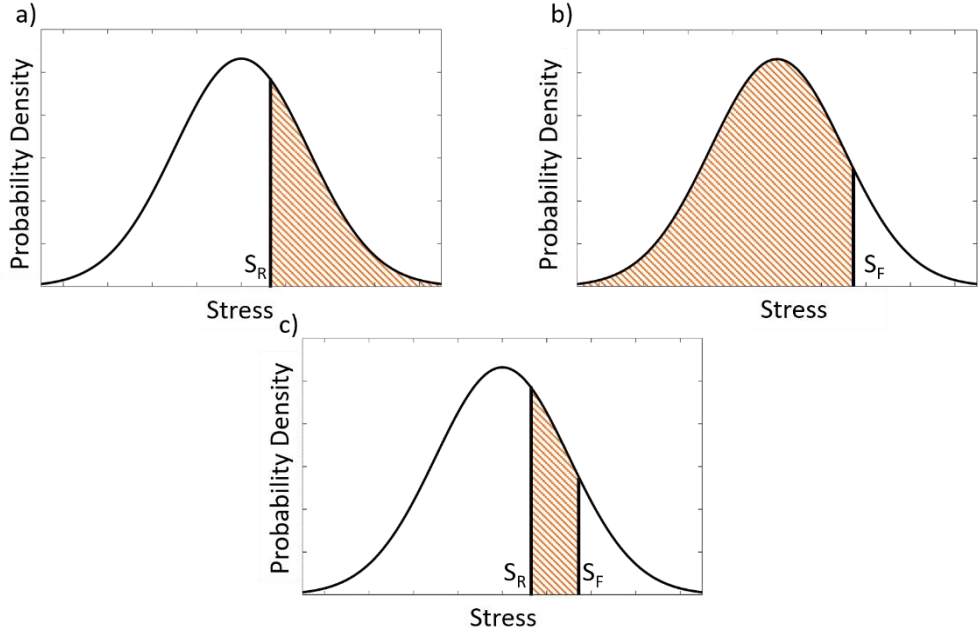

*Figure 1: Representation of probabilities of collecting results for fatigue strength tests. a) shows probability of a runout, implying that the specimen fatigue strength is not below $S_R$ – i.e. that it is right-censored. b) shows a failure at $S_F$ only, and c) a runout and failure pair.*

The likelihood $L_i$ of a single test $i$ is evaluated following the approach of Engler-Pinto et al [11,12] and Smalakys et al [46], represented graphically in Figure 1 and Eq. (4), where $\Phi(x)$ is the cumulative distribution function (CDF). The staircase test can either runout (where $j_i = 0$, and said as "right – censored") or fail (where $j_i = 1$, or it is "left – censored"). For step testing, the stress is incremented until failure, so there are no runouts. The specimen can either fail on the first step ($j_i = 1$, $k = 1$, left– censored) or fail on any other step ($j_i = 1$, $k \neq 1$, interval – censored). Therefore, Eq. (4) can be rewritten as Eq. (5):

$$L_i = p(j_i|[x_{i,k},\theta]) = \begin{cases} 1 - \Phi(x_i), & \text{Runout} \to \text{right} - \text{censored} \\ \Phi(x_i), & \text{Failure} \to \text{left} - \text{censored} \\ \Phi(x_{i,k}) - \Phi(x_{i,k-1}), & \text{Step} \to \text{interval} - \text{censored} \end{cases} \qquad (4)$$

$$L_i = p(j_i|[x_{i,k},\theta]) \qquad (5)$$
$$= \begin{cases} \Phi(x_i) \cdot \delta_{1,j_i} + [1 - \Phi(x_i)] \cdot \delta_{0,j_i} & - \text{Staircase} \\ \Phi(x_{i,k}) \cdot \delta_{1,k} \cdot \delta_{1,J_i} + [\Phi(x_{i,k}) - \Phi(x_{i,k-1})] \cdot (1 - \delta_{1,k}) \cdot \delta_{1,J_i} & - \text{Step} \end{cases}$$

Where it is assumed that the step test stress is incremented until failure. The likelihood $p(j|[x,\theta])$ with N specimens is obtained by multiplying the likelihoods for all specimens Eq. (6):

$$L = p(j|[x,\theta]) = \prod_{i=1}^{N} L_i \qquad (6)$$

## 2.1 APPLYING BAYESIAN MAXIMUM ENTROPY SAMPLING TO FATIGUE DATA

The goal of precisely estimating distribution parameters can be thought of as generating the most compact, or tightly constrained, joint posterior parameter distribution. To choose the next stress to test that will maximize this predictive power, Ryan, Drovandi et al [43] recommend maximising the expected gain in Shannon information $I$ Eq. (7) in the joint posterior. The next stress to test is therefore chosen to maximise the expected information as in Eq. (8) from sampling at a subsequent stress $x_{i+1}$. Often called entropy, this expected information can be intuited as a quantitative measure of the compactness of the posterior distribution, a derivation can be found in [42–44,47]:

$$I(p(\theta|[j_{i+1}, x_{i+1}x, j])) = -\ln(p(\theta|[j_{i+1}, x_{i+1}, x, j])) \qquad (7)$$

$$ent(p(\theta|[j_{i+1}, x_{i+1}, x, j])) = E[I(p(\theta|[j_{i+1}, x_{i+1}x, j]))]$$
$$= -\int d\theta \; p(\theta|[j_{i+1}, x_{i+1}x, j]) \ln(p(\theta|[j_{i+1}, x_{i+1}, x, j])) \qquad (8)$$

To maximise the compactness of the posterior distribution, we define a gain in entropy $\Upsilon(j_{i+1}|x_{i+1}, j, x)$ from the addition of a possible data $j_{i+1}$ (a failure or runout) at $x_{i+1}$:

$$\Upsilon(j_{i+1}|x_{i+1}, j, x) = ent(p(\theta)) - ent(p(\theta|[j_{i+1}, x_{i+1}, x, j])) \qquad (9)$$

As $ent(p(\theta))$ does not vary with data, we can neglect this term. We therefore build a utility function $U(x_{i+1})$ to evaluate the expected value in the change of entropy $E[\Upsilon(J|x_{i+1}, j, x)]$ of including the datapoint $x_{i+1}$ over all possible outcomes $J$ of the next specimen:

$$U(x_{i+1}) = -E[\Upsilon(J|x_{i+1}, j, x)]$$
$$= \sum_J p(J|[x_{i+1}, j, x], \theta) \int d\theta \; p(J|[x_{i+1}, j, x], \theta) \ln(p(J|[x_{i+1}, j, x], \theta)) \qquad (10)$$

This can be intuited as evaluating the expected gain of information for testing the next specimen at some stress, for all outcomes. The maximisation of this utility function therefore allows for the greatest expected entropy gain.

For staircase-type testing, the utility function $\mathbf{U}^{staircase}(x_{i+1})$ is built by considering both possible outcomes: a runout or failure at stress $x_{i+1}$ and evaluates this expected entropy change as a function of the test stress:

$$\mathbf{U}^{staircase}(x_{i+1}) = \Upsilon(0|x_{i+1}, j, x) \; p(0|[x_{i+1}, j, x], \theta)$$
$$+ \Upsilon(1|x_{i+1}, j, x) \; p(1|[x_{i+1}, j, x], \theta) \qquad (11)$$

This can be understood as determining the expected value for each possible outcome through the entropy gain, and its associated probability given the previous data. In this case, we aim to find the test stress $x_{i+1}$ which maximises $\mathbf{U}^{staircase}$.

For step-type testing, the design space allows for varying two test parameters: the starting stress $x_{i,0}$, as well as the step size $\xi$. The utility function $\mathbf{U}^{step}(x_{i+1,1}, \xi)$ is similarly built by considering possible outcomes as discussed above. Since the specimen could fail on any step – the expected $\Upsilon$ from each step contributes to the expected utility. Therefore, the utility function becomes Eq. (12):

$$\mathbf{U}^{step}(x_{i+1,1}, \xi) = \Upsilon(1|x_{i+1,1}, \boldsymbol{j}, \boldsymbol{x})\, p(1|[x_{i+1}, \boldsymbol{j}, \boldsymbol{x}], \boldsymbol{\theta}) \\ + \sum_{k=2}^{N_k} \Upsilon(1|x_{i+1,k}, \boldsymbol{j}, \boldsymbol{x})\, p(1|[x_{i+1,k}, \boldsymbol{j}, \boldsymbol{x}], \boldsymbol{\theta}) \tag{12}$$

where the first term denotes failing on the first step, and the sum across $k$ evaluates the utility of failing at each subsequent step, recalling that $x_{i+1,k} = x_{i+1,1} + k\xi$. Here, $N_k$ is the number of possible test stresses ($k$) from the starting stress to the maximum possible test stress. The sum need only be defined up to the point where the probability of collecting a runout and fail pair, $p(1|x_{i+1,k}, \boldsymbol{\theta})$, becomes infinitesimally small. In the computational implementation, as seen on GitHub under "BayesOptFatigue", this was arbitrarily taken as the maximum value of the location or mean parameter plus six times the maximum spread value, which was shown to be sufficient.

## 3 COMPUTATIONAL FRAMEWORK

To validate our framework, a simulated material's fatigue response is generated. We arbitrarily generalise the fatigue response for any life as a modified Basquin model with a fatigue limit, to determine the fatigue strength at a certain life Eq. (13):

$$S_{mean} = \alpha \log(N)^{-\beta} + c \tag{13}$$

where the Basquin constants $\alpha$, $\beta$, and c are taken in this simulated system to be 600 MPa/log(cycle), 1, and 300 MPa, respectively. As such, the mean fatigue strength at $10^6$ cycles is 400 MPa. The modelling of the entire fatigue curve is at this stage arbitrary, but this is useful when extending the Bayesian approach to multiple lives [48].

We use a Monte-Carlo based method for generating a simulated sample set with many fatigue responses by perturbing this mean response in stress by a distribution with some characteristic spread. These simulated results use a Normal distribution as in Eq. (3) which is chosen as a simple model, though readers are invited to the literature on the basis for fatigue strength models [49,50,59–62,51–58], and six other distributions are implemented in the code. Thus, a specimen with a determined fatigue response is a member of a sample set with a broad fatigue response. The simulation treats each sample curve as unique, and to be tested as in the experimental protocols (e.g. they can only fail once). This can be written as Eq. (14):

$$S_i = \alpha \log(N)^{-\beta} + c + \delta(0, \sigma) \tag{14}$$

where $\delta(0, \sigma)$ is a random number with distribution centred about zero, and a spread (e.g. standard deviation) defined by $\sigma$. This results in a mean fatigue response at $10^6$ cycles is 400 MPa, and a spread of $\sigma$. In this framework, a single specimen's curve is selected to be subject to the test protocol, for either a single test or multiple step tests. For the selected specimen, if the fatigue strength at life log $(N)$ is higher than the test stress, it has not failed, while if the fatigue strength is lower than the test stress, it has failed. This allows for the simulated testing of large sample sets and comparison between protocols (Figure 2).

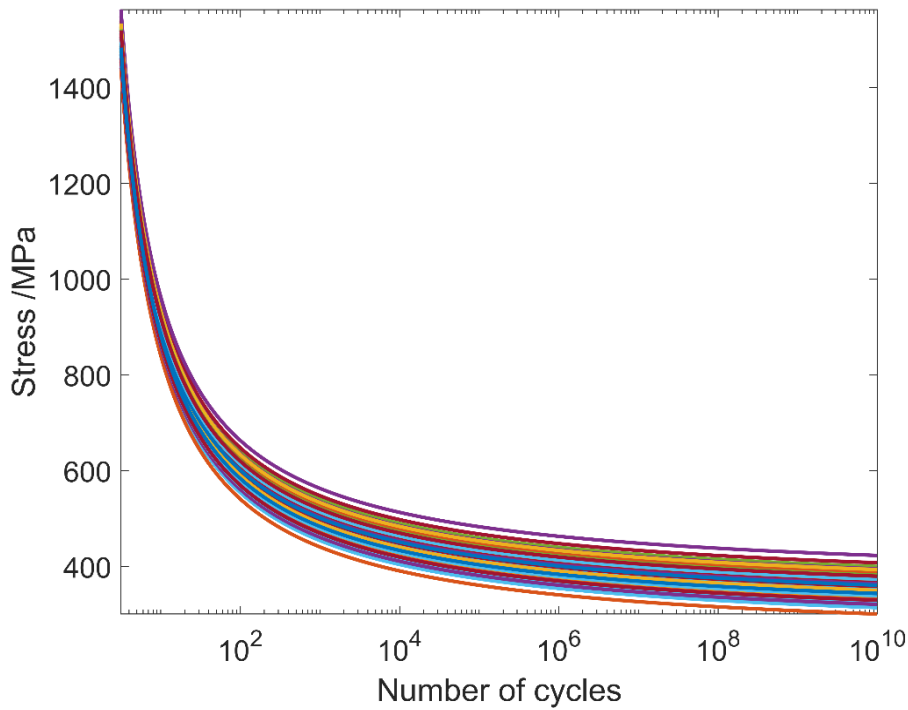

*Figure 2: Example simulated sample set with each specimen modelled as a Basquin curve perturbed in stress by a normally distributed random variable.*

In this computational method, the prior is built as a discrete N-Dimensional array with the characteristic distribution parameters along the axes such as location and spread, or mean and standard deviation in the case of a Normal distribution. The range of values of parameters defines the extent of the flat prior and the grid size affects the precision of model parameter estimates. The sequential evaluation and normalisation of the posterior is therefore Eq. (15):

$$p(\theta|x) = \frac{p(\theta|x_{1:i-1})p(x_i|\theta)}{\sum p(\theta|x_{1:i-1})p(x_i|\theta)} \qquad (15)$$

As such, each element in this array with coordinate of $(\theta_1, \theta_2 \ldots)$ gives the probability that these discrete parameter values represent the data. $ent(p)$ is therefore calculated by summing the probability contributions from every possible combination of model parameters of the desired distribution.

A flat prior is initialised by creating an array of equal values such that $\sum p(\theta|x) = 1$. A more detailed view of this process is included in the README of the GitHub repository [45].

## 3.1 BAYESIAN STAIRCASE PROTOCOL

The Bayesian Staircase collection strategy (Figure 3) broadly resembles the conventional staircase method in three ways: the interpretation of the algorithm's behaviour, an approximate similarity in collection in the initial points, and an analysis based on a sample set containing failed specimens and runout data.

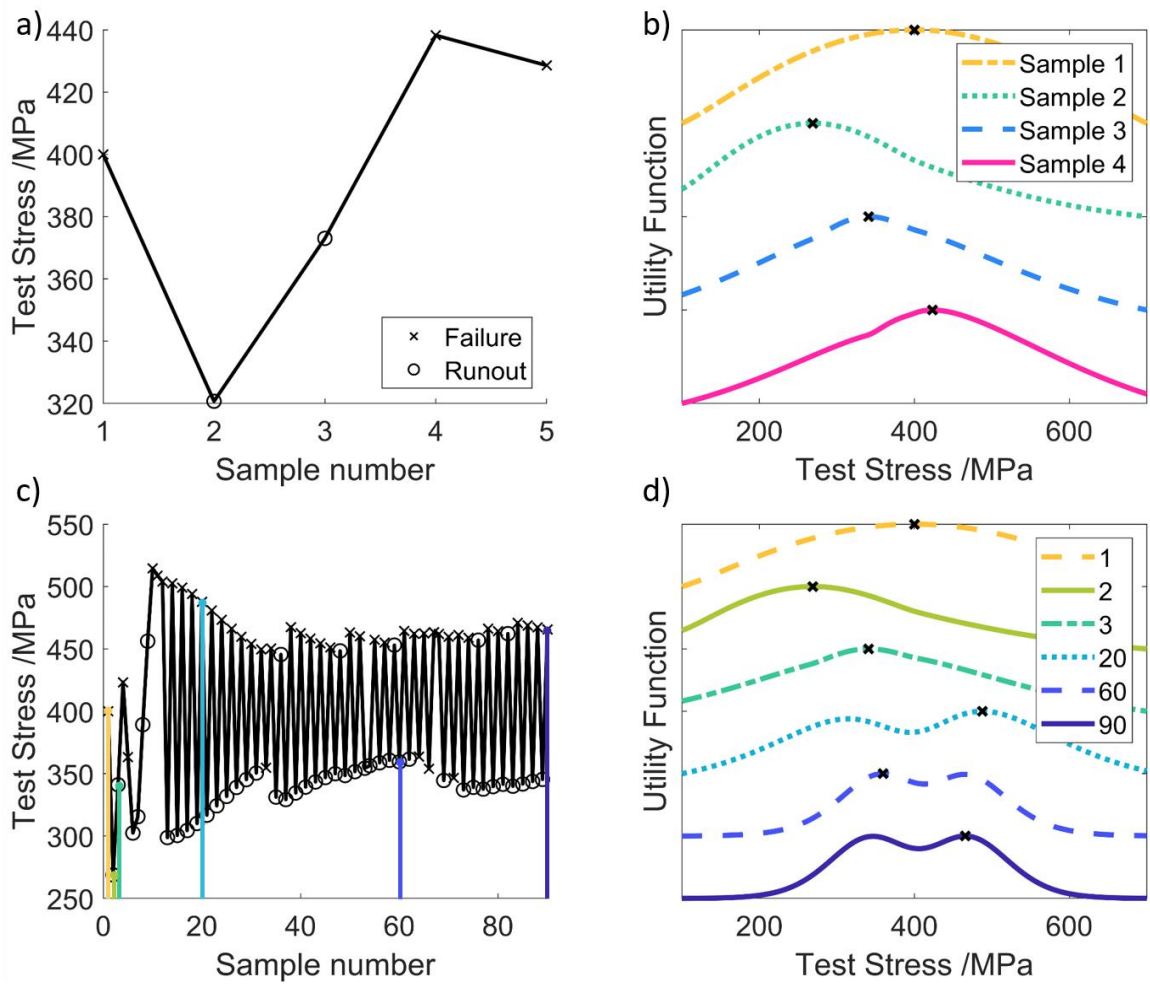

*Figure 3: Staircase plot (a, c) and associated utility function plots (b, d) for 4, and 90 specimens. The simulated sample set had a mean of 400 MPa and distribution width of 40 MPa.*

The conventional staircase protocol implicitly infers that the fatigue strength is below the test stress if a specimen has failed, and vice versa. The Bayesian protocol can be seen to follow this similar process, but considers all prior samples when determining the next test stress rather than following a rigid protocol. The first few specimens perform a binary search to estimate the mean fatigue strength. As the estimate becomes more precise, there is less information to be gained by testing at the centre of the distribution, and the algorithm then tries to determine the standard deviation. This can be seen in Figure 3) d) as the transition in the utility function to a bimodal curve with peaks at approximately one standard deviation from the mean. This transition quantifies the intuitive approach of estimating the mean followed by the spread of a sample set, similar in behaviour to Never's D-optimal test strategy [63].

The evolution of the estimate in the joint posterior can be seen in Figure 4. The Bayesian protocol can be seen to constrain the estimates for mean and standard deviation, with fewer model parameters falling within the 90% HPD surface as data are collected. It can likewise be seen that the estimate for mean is determined within 10 samples to a reasonable degree of confidence, while the standard deviation requires further testing.

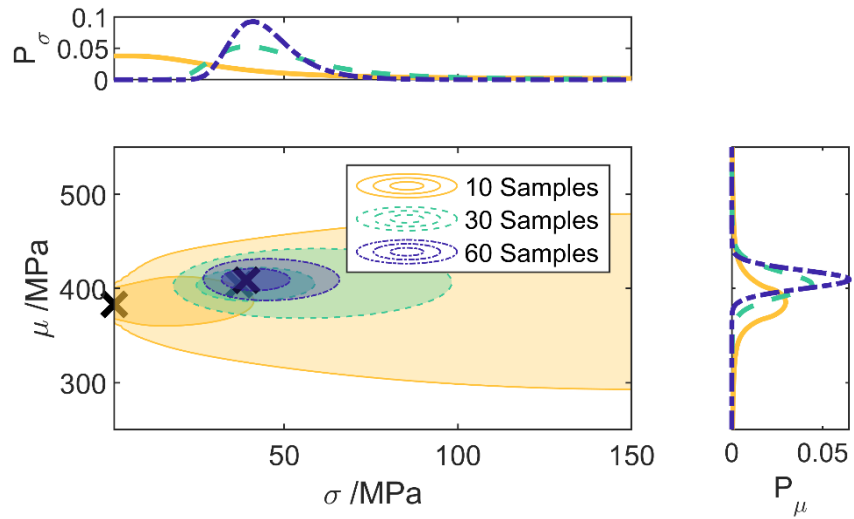

*Figure 4: Evolution of joint posterior surface for Bayesian Staircase protocol. The simulated sample set had a mean of 400 MPa and a distribution width of 40 MPa. Contours are at 50 and 90% HPD interval. The 2D array of model parameters can be seen, with projections showing the probability for each individual parameter.*

The Bayesian protocol has several advantages to the regular staircase method: very little information is required about the sample set prior to testing, testing parameters are not reliant on prior knowledge, and a faster convergence is obtained. The first two points follow from the protocol's process, and the latter can be seen in Figure 5. This figure shows the trend in error as a function of specimen number for various protocol parameters, as well as the error after 100 samples as a function of the ratio of stress increment to the true distribution standard deviation. The percentage error is calculated relative to the known true distribution parameters, for example for the mean: $Error_\mu = 100 * \frac{\mu_{estimate} - \mu_{real}}{\mu_{real}}$ .

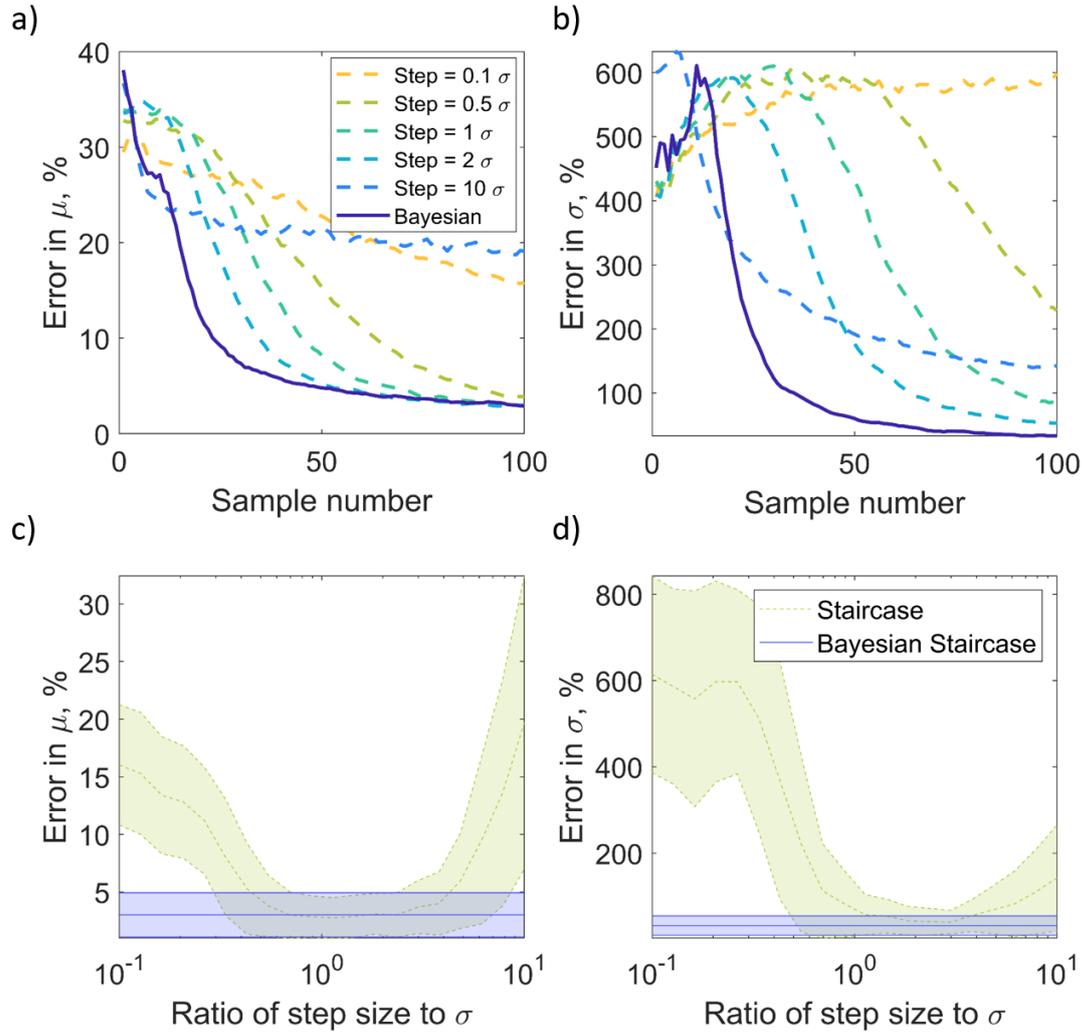

*Figure 5: Convergence plot of a) mean and b) standard deviation for a variety of step sizes, and comparison of error after 100 samples in c) mean and d) standard deviation between Staircase (red) and Bayesian Staircase (green) protocol. The simulated sample set had a mean of 400 MPa and a distribution with of 40 MPa, and the simulations in c) and d) are repeated 200 times to generate a standard deviation in the values (shaded regions). The comparison plot in c) and d) can be seen to be a cross section of a) and b) after 100 samples for a larger number of step ratios. a) and b) lines are smoothed in the x axis with a window of 5 samples.*

The behaviour of the conventional method can be seen to be strongly sensitive to the protocol parameters. This implies that the choice of step size greatly influences the error in the result for conventional staircase testing. The Bayesian protocol outperforms or matches the conventional protocol for all possible step sizes.

## 3.2 BAYESIAN STEP PROTOCOL

We create a protocol that finds an optimal step size and starting stress, setting out the stress levels for an individual specimen. For constant life testing, the optimum step size is always the smallest step size as it provides the smallest interval of fatigue strength for the specimen. This trivial result reflects the trade-off described above, but without regard for damage effects. Therefore, the benefit of this Bayesian approach in comparison to the conventional protocol arises when using larger step sizes: where damage is less of a concern and the role of starting stress can greatly influence accuracy. Thus, the dimensionality of the utility surface can be reduced to only considering varying the starting stress with a constant step size. Here, we choose 160 MPa, or 4 times the standard deviation, shown in Figure 6. This significantly larger step size is used only as a demonstrator of the effect of starting stress.

Generally, the protocol follows similar behaviour to the Bayesian Step protocol. The starting stress has large variations at low numbers of samples where varying the possible failure stresses gives a quicker estimation of the mean fatigue strength. When the step size is larger than the width of the sample strength distribution, the starting stress tends towards where the stress levels intersect stresses approximately one standard deviation away from the mean, analogous to the behaviour noted in section 3.1. Conversely, when the step size is much smaller than the width of the sample strength distribution, there is comparatively less to gain by varying starting stresses.

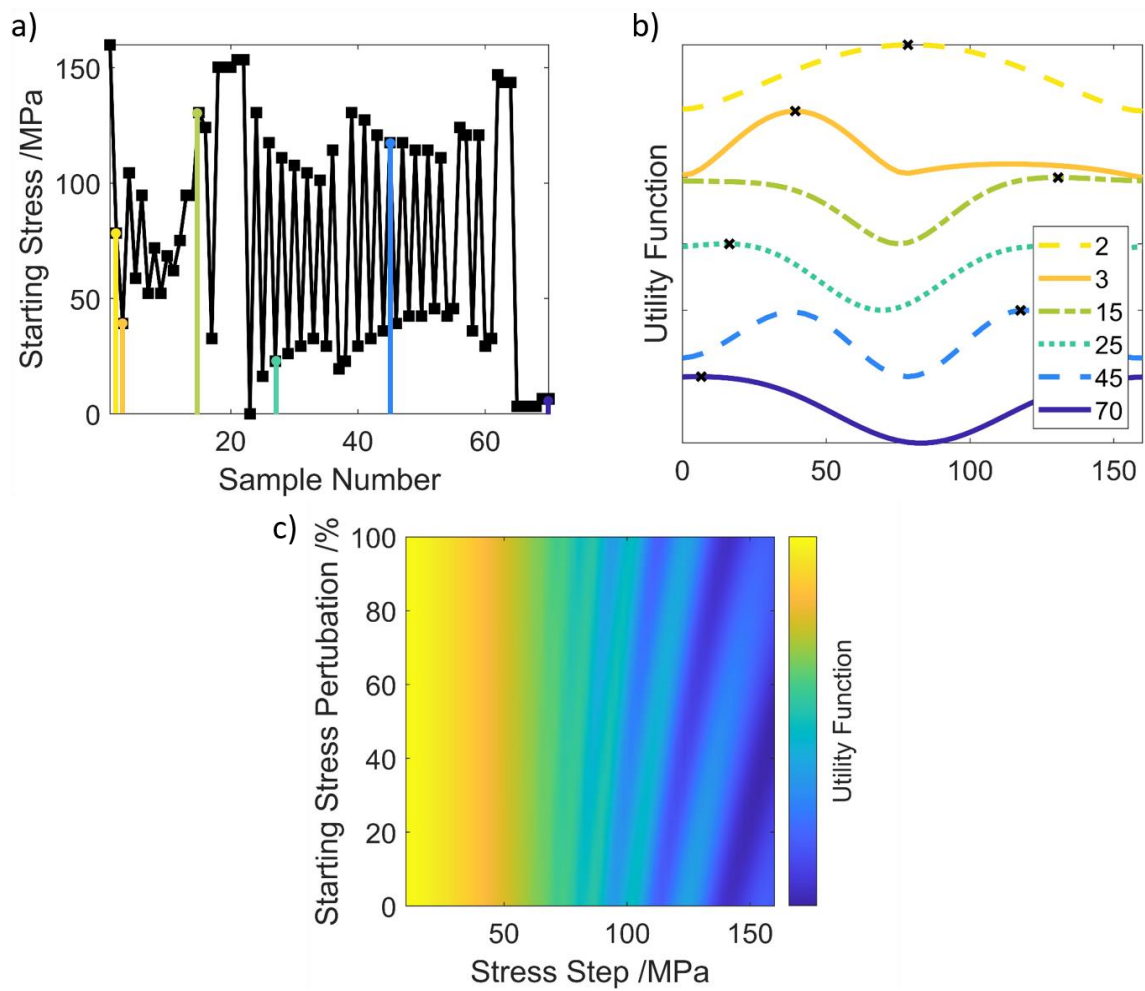

*Figure 6: a) Staircase plot of the Bayesian Stress Step protocol, showing the starting stress as a function of specimen number. The specimens simulated had a mean strength of 400MPa and a distribution width of 40 MPa. The stress step was 160 MPa, with a maximum starting stress of 160 MPa. b) Utility function showing the relative utility of various starting stresses, at a selection of specimen numbers, colour-coded as in a). c) Utility at 70 samples is shown when allowing for a variation in step size, showing the preference for smaller step sizes. In this case, the vertical slice at a stress step of 160 MPa is equal to the curve at 70 specimens in b).*

As in the case of the staircase protocols, a comparison can be made between the conventional stress step protocol and its Bayesian counterpart, as a function of the stress increment size shown in Figure

7. As discussed in section 8.2, mean values of error for step testing do not capture the bi-modal distribution in error, i.e. an experimentalist is most likely to observe higher error for small numbers of samples. It should also be noted that the error level, even at this reduced number of specimens, is lower than for staircase testing at small step sizes. This is in line with common practice, highlighting the trade-off in choosing protocols between efficiency and concern about damage.

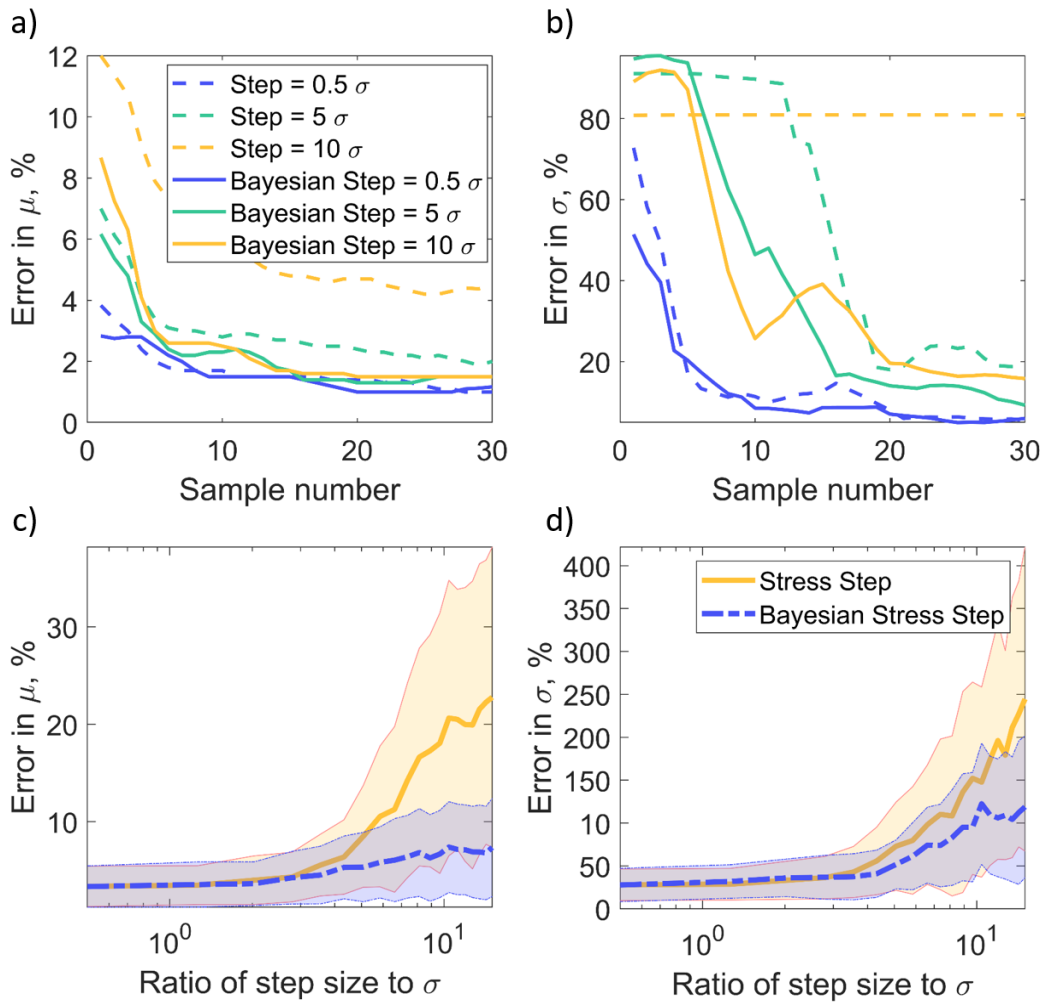

*Figure 7: Convergence plot of a) mean and b) standard deviation for a variety of step sizes, and comparison of error after 30 samples in c) mean and d) standard deviation between conventional (red) and Bayesian (green) for a variety of ratios between step size and true distribution width. The simulated sample set had a mean of 400 MPa and a distribution with of 40 MPa, and the simulations in c) and d) are repeated 200 to generate a standard deviation in the values (shaded regions). The comparison plot in c) and d) can be seen to be a cross section of a) and b) after 30 samples for a larger number of step ratios. a) and b) lines are smoothed in the x axis with a window of 5 samples.*

The benefit from the Bayesian Step protocol comes when using larger step sizes and the relative difference of starting a fraction of a step away (Figure 7). For these larger step sizes, the Bayesian protocol has a significantly better performance in estimating both mean and standard deviation. For example, the Bayesian protocol with a 15 σ step size and obtained a better estimate of mean than a conventional step protocol with 5 σ step after 30 samples.

Stress step tests can therefore be used with comparatively less concern over damage by using significantly larger step sizes and varying the starting stress. Compared to the conventional protocol,

## 3.3 STATISTICAL MODEL SELECTION

The choice of statistical distribution model to describe fatigue strength will strongly affect the probability of failure especially away from the median stress. This has motivated work in determining model suitability [49,50,59–61,51–58]. We provide a tool for the comparison of different statistical models given experimental data collected, such that model suitability can be explored.

This work focusses on Bayes Information Criterion (BIC), shown in Eq. (16). BIC is based on the maximum log-likelihood $\hat{L}$, number of observations $n$, and the number of parameters $k$ in the model. The term with number of parameters penalizes against models with more parameters as they will be more liable to overfitting. Models with a lower BIC should be selected over other models.

$$BIC = k \ln(n) - 2 \ln(\hat{L}) \qquad (16)$$

$$\hat{L} = \max p(\boldsymbol{j}|[\boldsymbol{x}, \boldsymbol{\theta}]) \qquad (17)$$

$\hat{L}$ is the maximum value of the likelihood function Eq. (17) which describes the probability of observing data given a model $M$, data $\boldsymbol{j}|\boldsymbol{x}$, and parameters $\boldsymbol{\theta}$. Thus, $\hat{L}$ gives a measure of the goodness of fit of the data to the best model of that functional form. Following Kass and Raftery [64], the value of the BIC can give evidence for a model over another model with lower values indicating more evidence. To compare two models, $\Delta BIC$ can be defined as $\Delta BIC = BIC_1 - BIC_2$, and shows evidence for $BIC_1$ if it is negative. Values of $\Delta BIC$ between 0 and 2 indicates weak evidence, 2 and 6 indicates substantial evidence, and 6 to 10 indicates strong evidence for the lower BIC.

A simulated experiment was performed to observe the effectiveness of conventional and Bayesian protocols in distinguishing different distribution models, with the result shown in in Figure 8. The ability of this criterion to select from these distributions can be seen to be dependent on the protocol used as well as the model of the input sample distribution: step-type protocols resolve the distribution type much more effectively than staircase-type protocols.Figure 8 Repeat experiments comparing to 3-parameter Weibull, Gumbell, and Frechet produced similar distinction between distribution models, with stress-step methods providing stronger distinction between distributions. More generally, step-type protocols with larger step size relative to distribution were more effective at distinguishing distribution type, and it is harder to distinguish Log-Normal from Normal.

However, it must be noted that the distinction between these distributions remains unclear after 80 specimens, and further testing is required to develop a confident distinction with these protocols. This does suggest the possibility of developing a utility function with an experimental aim to maximise the distinction between distributions.

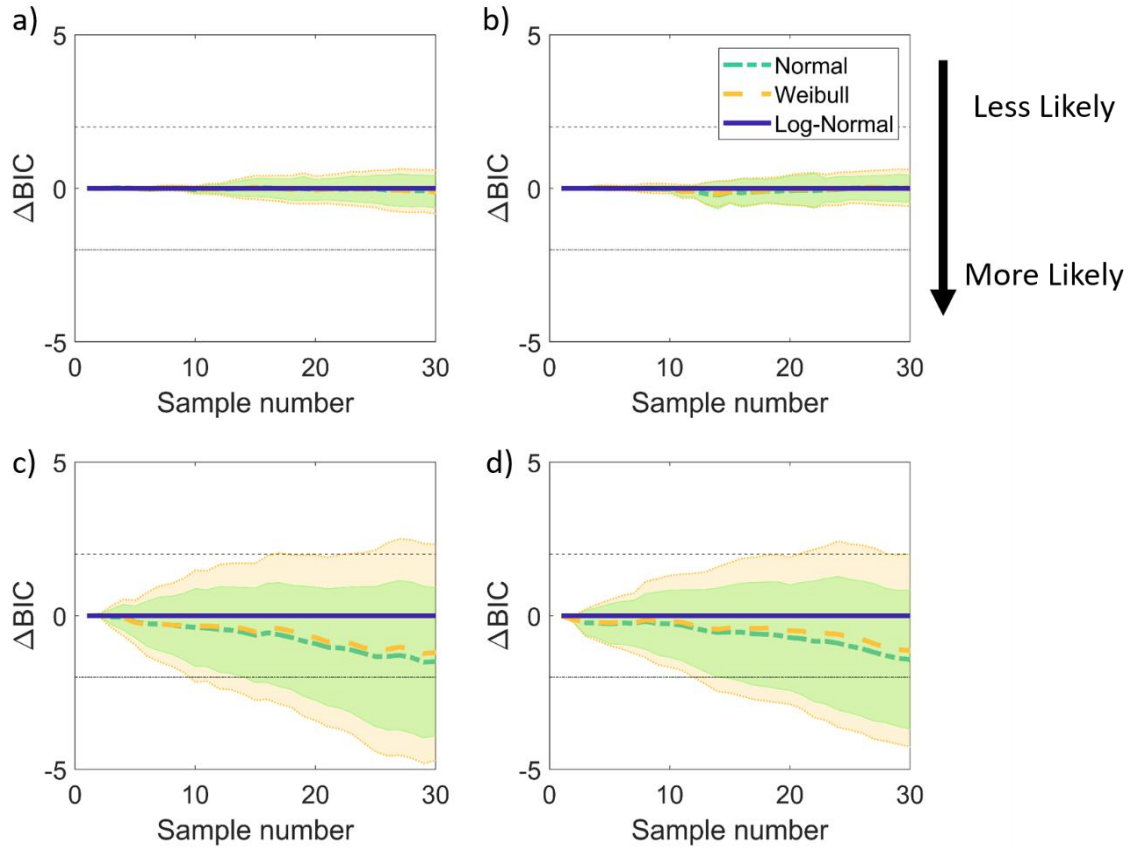

*Figure 8: Evolution in ΔBIC (difference in Bayes Information Criterion) for data with Normal distribution and mean of 400 MPa with standard deviation 40 MPa. ΔBIC is calculated relative to BIC for 2-Parameter Weibull distributions, with the lowest BIC indicating a distribution with stronger evidence. a) Staircase, b) Bayesian Staircase, c) Stress Step, d) Bayesian Stress Step protocols based on 100 repeats (shaded region corresponding to one standard deviation). Step size was 40 MPa and starting stress was 380 MPa for Staircase and 120 MPa for Step. The dashed lines represent the level of evidence shown as in Kass and Raftery [64].*

## 3.4 COMPARING PROTOCOLS AND LIMITATIONS

The comparison between conventional and Bayesian-type protocols as shown in Figure 5 and Figure 7 are a result of our definition of convergence. However, this visualisation significantly reduces the information present in the simulations, and our definition of convergence is just one of many ways to compare results. The current definition only considers the deviation of the MAP from the true value. Further considerations can be made, for instance, by observing the gradient at which the MAP reaches a value and does not change, the number of specimens required for this converged value not to change by some threshold, and analysis considering the distribution in parameter probabilities. However, regardless of the definition, it was observed that Bayesian counterparts to conventional methods lead to faster, narrower, and more accurate estimates of both parameters, independent of prior knowledge. Furthermore, in comparing the protocols considered in this analysis, there is a clear difference between the rate of convergence between single-stress tests and multiple-stress tests. Stress step methods may be favourable when fewer specimens are available, whereas single-stress tests may be favourable when conducting tests in a shorter amount of time or without concern of loading history.

Examining further, these results also give an insight into the common practice of the conventional protocols. The staircase method can be seen to have the lowest error in determining the mean when the stress increment is near the standard deviation, reflecting the common practice discussed as early as in the original Dixon-Mood reports [3]. However, it can also be seen that this common practice forgoes the nuance in this decision. Figure 9 shows the convergence plot as shown in Figure 5 with only the staircase results after 100 samples, overlaid with the errors as obtained after 30 and 60 samples.

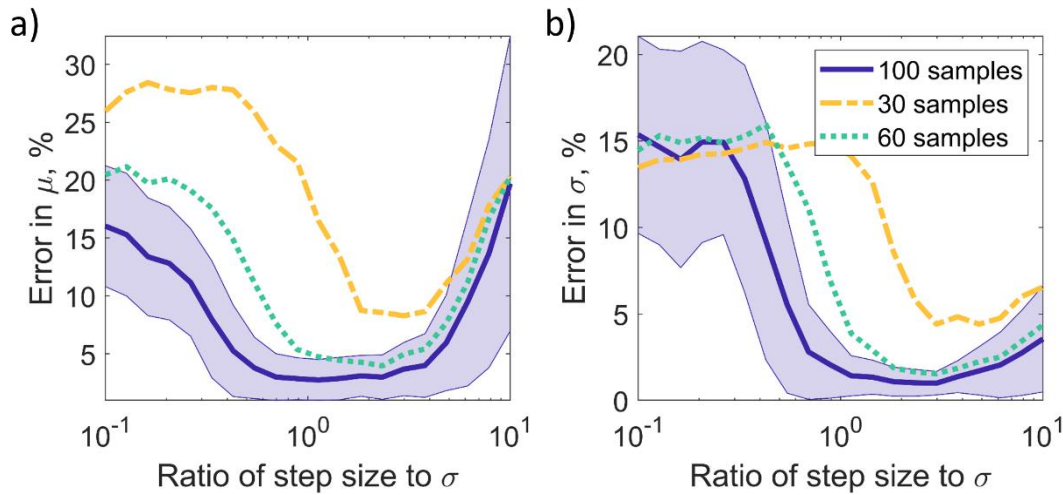

*Figure 9: Comparison of error in a) mean and b) standard deviation as shown in Figure 5 c) and d) examining only the staircase protocol for a variety of sample numbers.*

While the optimal increment for measuring the mean to the lowest error is indeed equal to the standard deviation, this is not true for measuring the standard deviation. Furthermore, the optimal increment for measuring both mean and standard deviation changes with sample number. Initially, it is more effective to collect data with a larger increment, a result that is reflected in the Bayesian protocol during the binary search and ensuing collection. This highlights the significant advantage of the Bayesian protocol: an evolving sequential adaptation of the testing parameters to the data.

However, whilst the Bayesian methods avoid requiring estimates of the fatigue system to be measured in determining starting stress or step size, the construction of the prior itself is necessary. Both the range in parameters values as well as the coarseness of this discrete mesh require some broad estimation, though this requirement is substantially more flexible: the binary search is sufficiently quick at determining the mean if it is within the extent of the prior. Coarseness of the discrete mesh is only a concern when, for reasons of computational efficiency, the mesh has been coarsened to such a degree that the model predicts likelihood for one set of parameters, and zero probability for neighbouring parameters. During collection of either simulated or experimental results, this can be easily visualised and amended.

This method lends itself to generalisation to allow for a broader set of experimental goals. The method can be generalised to extend across the P-SN space [48]. From earlier discussions of distribution types, it may be possible to generate a utility function that maximises the distinction between distribution types [44]. This represents a potential for future work, highlighting the need in engineering practice to ensure predictive capability at the extremes of the distribution.

# 4 EXPERIMENTAL VALIDATION

An experimental validation of the above simulations was carried out to confirm the basis of the Monte Carlo sample set and demonstrate the protocols in a physical laboratory setting.

Fatigue data is generated using a novel small scale ultrasonic fatigue setup developed by J Gong. Data are collected at 20 kHz to rapidly generate SN data on a laser cut foil of 304L stainless steel. Figure 10 shows the unique geometry, with the exact fatigue response not being used to infer any property of the base material.

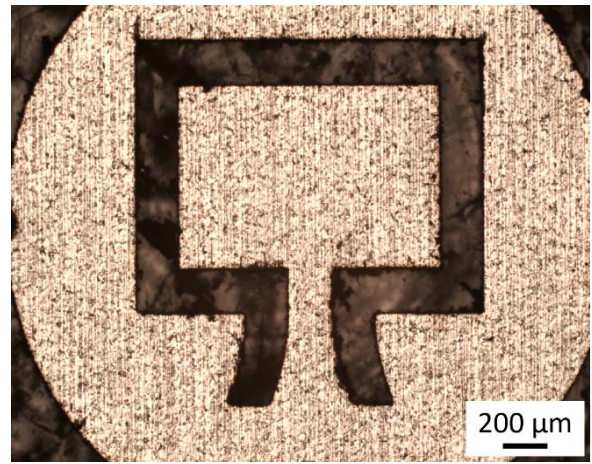

*Figure 10: Optical micrograph of specimen geometry. The length and width of the cantilever are 600 and 200 µm, and the foil from which it is cut is 100 µm thick. The system acts as a tuned harmonic oscillator with the large rectangular mass acting as a counterweight. The geometry is designed in Finite Element Analysis (FEA) to determine the stress during vibration.*

The specimens were randomly extracted for positions across the foil of material and assigned to protocols in a randomised order to reduce time-dependent effects. Conventional protocol parameters used are summarised in Table 3, and were based on best estimates from previous testing on similar specimens with the same method: a mean of approximately 300 MPa and a spread of ~30 MPa. It should be noted that this guidance from preliminary testing is a requirement to set up conventional testing but is not needed for the Bayesian protocols.

*Table 3: Experimental Protocol Parameters*

| Protocol | Starting Stress (MPa) | Stress Step (MPa) |
|---|---|---|
| *Staircase* | 325 | 30 |
| *Step* | 150 | 30 |

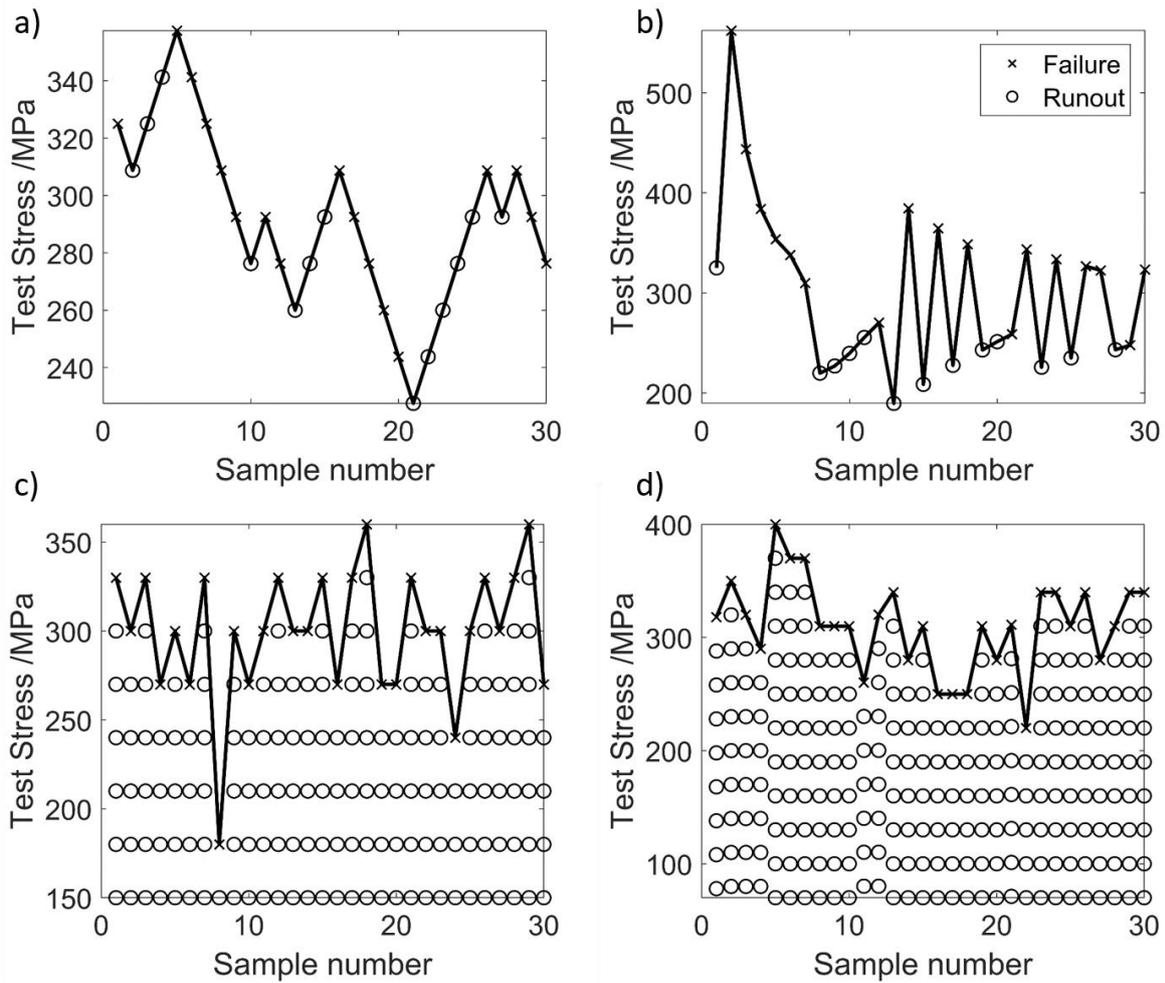

*Figure 11: Staircase Plots for the four experimental datasets. a) Regular staircase test, b) Bayesian Staircase, c) Regular Stress Step, d) Bayesian Stress Step. The stress increment in all tests (except Bayesian Staircase) is 30MPa, and the starting stress for the Stress Step protocol is 150MPa.*

Figure 11 shows the raw data from each of the protocols. All protocols follow the expected behaviour, including a similarity to the Bayesian behaviour as shown in section 3.

When analysing the results of all 120 specimens together – representing the most precise estimate of the fatigue behaviour - the material system has a fatigue strength of 290 ± 10 MPa at $10^6$ cycles, with a spread of approximately 40 ± 10 MPa. It can be seen in Table 4 and Figure 12 that the Bayesian protocol equivalents outperformed their conventional methods, with more accurate MAPs and narrower HPD intervals, despite good initial assumptions for starting stress and stress increment size in the conventional protocols. The analysis of these results via conventional methods, such as Dixon-Mood for staircase, or fitted distributions for stress steps, are included alongside their respective Bayesian estimates in Table 4.

*Table 4: Results for the experimental campaign to compare the four sampling protocols.*

| Protocol | Conventional Analysis, MPa | | Bayesian Analysis, MPa: MAP, [95% HPD interval] | |
|---|---|---|---|---|
| | **Mean** | **Standard Deviation** | **Mean** | **Standard Deviation** |
| *Staircase* | Dixon-Mood analysis. N.B, the qualification factor is 0.18 (>0.3 required) | | 277 [113, 368] | 83 [33,300] |
| | 290 | 10 | | |
| *Bayesian Staircase* | | | 276 [233, 319] | 35 [17, 114] |
| *Step* | Fitted Normal distribution through regression: $R^2$ = 0.39. Value ± 95% confidence interval: | | 285 [268, 302] | 35 [26,51] |
| | 285 ± 15 | 37 ± 8 | | |
| *Bayesian Step* | | | 296 [277,315] | 39 [29, 56] |
| *Whole Dataset* | | | 288 [277,298] | 39 [32, 49] |

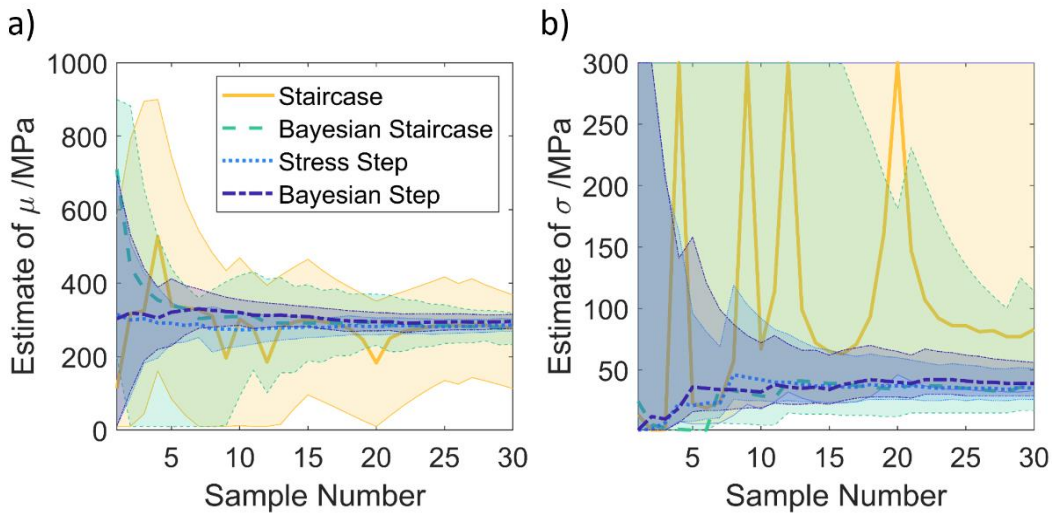

*Figure 12: Evolution in parameter estimates as a function of sample number, for a) mean and b) standard deviation for the four protocols tested. Line represents the MAP, and shaded regions indicate the 95% HPD interval.*

The conventional methods are unable to determine with reasonable confidence the parameter values that are best determined by considering the complete dataset – both the Dixon-Mood Qualification factor and the fitted distribution $R^2$ being too low. In addition, the Bayesian analysis of conventional data is better at generating reasonable estimates of parameters with a quantified error, with values within the range of the whole dataset estimate.

In general, the stress step methods unsurprisingly provide the tightest estimates of parameters. Ideally, a larger step size would have been used for the experiments to create more distinction between the Bayesian and conventional protocols, and would have minimised the effects of damage or coaxing that may have affected the result. However, in this experiment, no appreciable difference was seen in model parameters between step and staircase results.

Finally, a comparison can be made between distribution models. Accounting for all the tested specimens, there is more evidence for the 2 parameter Weibull distribution over all other common distributions as shown in Figure 13.

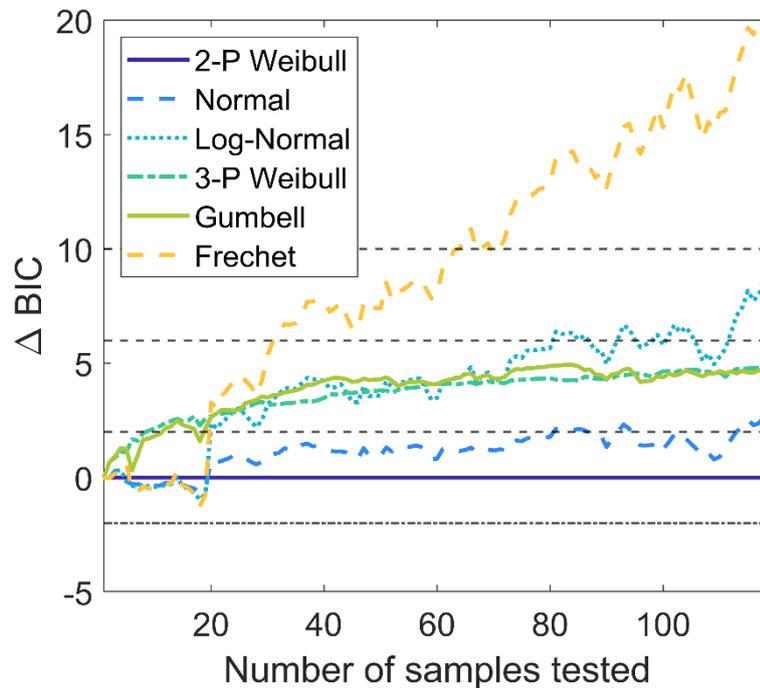

*Figure 13: Experimental evolution of ΔBIC combining data collected using all protocols, showing positive evidence for the Weibull distribution over normal and strong over log-normal [64]. Specimen 5 of the Bayesian step method results (Figure 5 d) was omitted due to experimental error. See Figure 8 for more details.*

## 5 CONCLUSIONS

We show that Bayesian statistics can be applied to fatigue strength testing. We present two open source alternatives to the conventional staircase and stress step protocols. These can be shown to provide better estimates for model parameters such as location and spread, such as mean and standard deviation. Furthermore, these methods are independent of rigid definitions for step size or starting stress, as required in conventional protocols. Using a Monte Carlo simulation framework, we demonstrate the benefit that these protocols bring, and validate these through experiment. Furthermore, simulation show insight into the behaviour underlying the conventional protocols, further demonstrating the limitations of fixed step sizes.

Finally, this framework permits comparisons of distribution models, allowing design engineers to make better trade-offs for weight and reliability. We also show with experimental data that a 2-parameter Weibull distribution may better fit fatigue strength data, and offer the reader the tool to compare distribution models for other fatigue data.

These tools can be applied to any pass-fail testing application, fatigue or otherwise, and is material independent. The method, built for both the collection of data during experiments as well as the analysis of collected data, is made freely available under the GitHub repository "BayesOptFatigue".

# 6 CRediT Taxonomy

CM Magazzeni: Conceptualisation, Data curation, Software, Formal Analysis, Validation, Investigation, visualisation, methodology, Writing – original draft, project administration

R Rose: Conceptualisation, Data curation, Software, Formal Analysis, Validation, Investigation, visualisation, methodology, Writing – original draft, project administration

C Gearhart: Conceptualisation, Resources, Software, Formal analysis, supervision, methodology, writing – review & editing

J Gong: Investigation, methodology

AJ Wilkinson: Funding acquisition, Writing – review & editing

# 7 Acknowledgments

CMM wishes to acknowledge Rolls Royce plc, the Royal Commission for the Exhibition of 1851, and the EPSRC for financial support. This work was authored in part by the National Renewable Energy Laboratory, operated by Alliance for Sustainable Energy, LLC, for the U.S. Department of Energy (DOE) under Contract No. DE-AC36-08GO28308. The views expressed in the article do not necessarily represent the views of the DOE or the U.S. Government. The U.S. Government retains and the publisher, by accepting the article for publication, acknowledges that the U.S. Government retains a nonexclusive, paid-up, irrevocable, worldwide license to publish or reproduce the published form of this work, or allow others to do so, for U.S. Government purposes.

# 8 APPENDIX

## 8.1 CHOICE OF INITIAL PRIOR

When choosing an initial prior for inference based on the normal distribution, two non-informative priors were investigated. As suggested by Box and Tiao [65] the Jeffreys prior Eq. (18) promotes models with small standard deviations.

$$P(\theta|x) \propto \sigma^{-1} \quad (18)$$

Alternatively, a constant prior over a range is used, which has the advantage that it is mathematically equivalent to a maximum likelihood approach. The choice of using a flat prior, as opposed to the Jeffreys prior, is justified by both its simplicity as well a notable feature in the trend of Shannon information at low specimen number [65,66]. When using a Jeffreys prior the Shannon information drops significantly after approximately 10 samples (Figure 14). This is generally observed as an initial under-estimation of the standard deviation, in line with the prior construction, followed by an increase in the estimate and uncertainty in the standard deviation. This overconfidence in a low standard deviation was deemed, particularly in the context of fatigue testing, as sub-optimal, despite the subsequent faster convergence to the same parameter values. For these reasons, a fully flat prior was used in the testing and simulation reported in this work.

However, it must be clear that a flat prior is also not assumption-free. Rather, the standard deviation is deemed equally likely to lie within the extent that is specified in the permitted 2D parameter space.

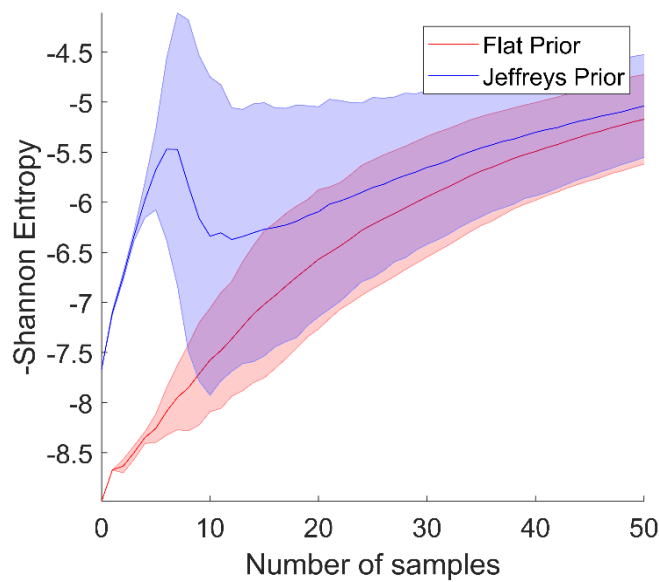

*Figure 14: Comparison of negative Shannon Entropy between tests with priors initialised as: uniform probability (red), and non-informative (blue). The tests were performed on a simulated set with strength 400 MPa and distribution width 40 MPa at $10^6$ cycles for 100 specimens using the Bayesian Staircase protocol. The shaded error bars represent the standard deviation between 500 repeats.*

## 8.2 Variability In Step Testing With Large Step Sizes

As discussed in the main text, step sizes in step testing should be as large as possible to minimise the effects of coaxing and damage on the fatigue failure distribution observed from a test campaign. However, the larger the step size, the higher the chance that the failure stress is "jumped over" during testing: specimens always survive and fail at the same stress levels. 1515

This can also be understood as placing a large importance on the initial test stress: if only few of the subsequent test levels intersect the distribution, it is easier to introduce bias. This bias produces a bimodal distribution in the MAP estimate with the modal value sometimes far away from the mean value (Figure 165), making the mean error for step testing not representative of what an experiment campaign is likely to observe.

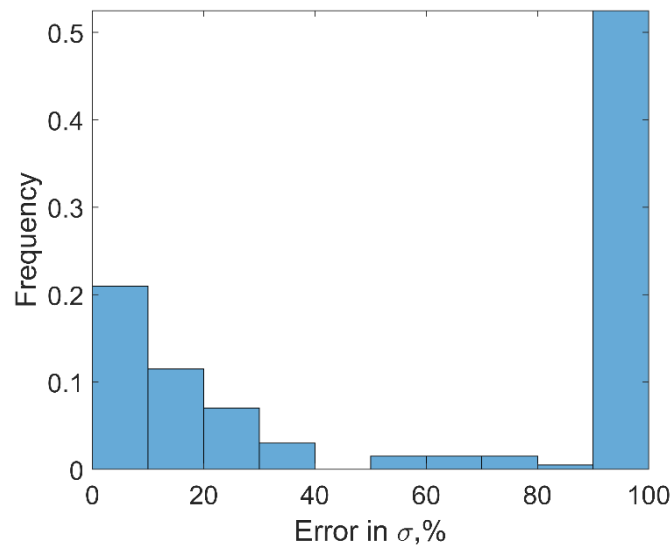

*Figure 165: Histogram of errors in conventional step testing after 5 samples, with a step size 5 times the true standard deviation of the simulation dataset, at 200 repeats with a random starting stress between 0 MPa and the step size.*